\documentclass[%
 reprint,
superscriptaddress,
 amsmath,amssymb,
 aps,
]{revtex4-1}

\usepackage{booktabs}
\usepackage{multirow}

\usepackage{subcaption}

\usepackage{bm} 
\usepackage{graphicx}
\usepackage{dcolumn}
\usepackage{bm}
\usepackage{hyperref}

\begin{document}

\preprint{APS/123-QED}

\title{Short-Hair Black Holes and the Strong Cosmic Censorship Conjecture}

\author{Zhiqin Tu}

\author{Meirong Tang}

\author{Zhaoyi Xu}%
\email{zyxu@gzu.edu.cn(Corresponding author)}
\affiliation{%
 College of Physics,Guizhou University,Guiyang,550025,China
}%


\begin{abstract}
The singularity problem has always been a focus of physicists' research. In order to solve this problem, Penrose proposed the cosmological censorship conjecture, but verifying this conjecture in different situations still faces many challenges. In the context of short-hair black holes research, this paper explores whether the Strong Cosmic Censorship Conjecture(SCCC) is obeyed by the universe when it is disturbed by a scalar field. In this paper, we explore whether the short-hair black hole satisfies the SCCC under scalar field perturbations. Using the Weak Gravity Conjecture (WGC) and the WKB approximation method within the framework of general relativity, we systematically analyze the behavior of short-hair black holes under different parameter conditions. The focus is on whether violations of the SCCC occur when the black hole approaches extremal conditions. The results show that when the charge \( Q \) of the short-hair black hole approaches its extremal value, the SCCC is violated. However, as the order \( k \) of the black hole’s metric equation and the angular momentum quantum number \( l \) increase, the phenomenon of SCCC violation is delayed. These findings indicate that the proximity of the black hole’s charge \( Q \) to extremality, as well as the values of the angular momentum quantum number \( l \) and the order \( k \), play crucial roles in exploring black hole physics and verifying the SCCC. This research not only reveals the behavior of short-hair black holes under extreme conditions but also provides a new perspective for further investigation of the SCCC.

\end{abstract}

\maketitle


\section{\label{sec:level1}Introduction}
General relativity predicts the existence of black holes, and the discovery of black holes further validates the correctness of general relativity under extreme gravitational fields, deepening our understanding of spacetime structure and gravitational phenomena. A black hole is a celestial object with extremely strong gravity, so strong that not even light can escape its gravitational pull. It can be described by Einstein's field equations and is usually formed from the gravitational collapse of a massive star at the end of its life. This collapse leads to extreme curvature of spacetime and generates a boundary called the event horizon, beyond which no matter or information can escape. In recent years, the detection of gravitational waves has further confirmed the existence of black holes. For example, the gravitational waves detected in 2015 by the LIGO and Virgo\cite{abbott2016observation} collaboration provided direct evidence of black holes. Furthermore, in 2019, the Event Horizon Telescope (EHT) successfully captured the shadow image of the supermassive black hole at the center of the M87 galaxy\cite{collaboration2019first}. These observational phenomena provide strong support for the correctness of general relativity's predictions in strong gravitational fields.

Although general relativity is highly successful in describing the external characteristics of black holes, it faces significant challenges when addressing internal black hole problems, particularly regarding the singularity. When a massive star exhausts its internal fuel, the core undergoes gravitational collapse, compressing into an extremely small volume, which leads to an extraordinarily strong gravitational field. This eventually forms a point of extremely high density and nearly zero volume, known as a singularity. The formation of the singularity marks an extreme deformation of spacetime. In such a scenario, general relativity and other classical physics theories lose their predictive power. The singularity theorems by Penrose and Hawking\cite{senovilla20151965}\cite{kunzinger2022hawking} demonstrate that under certain conditions, the formation of singularities is inevitable. However, general relativity cannot provide an effective physical description at the singularity, which limits its predictive power under extreme conditions. The existence of a naked singularity would pose a major challenge to general relativity, as it would lack an event horizon to enclose it, potentially allowing its gravitational effects to directly influence the external universe, leading to a breakdown in the causal structure of spacetime\cite{gielen2024black}. Therefore, the singularity problem of black holes reveals the limitations of general relativity, indicating the need for new theories to supplement or replace it, particularly under extreme gravitational conditions. To address this issue, Penrose proposed the cosmic censorship conjecture\cite{penrose1969gravitational}\cite{penrose1979singularities}, which asserts that naked singularities should not appear in the observable universe to ensure the completeness of spacetime and the stability of the causal structure.
\setlength{\parskip}{0pt}

Penrose's cosmic censorship conjecture exists in two main forms, the Weak Cosmic Censorship Conjecture (WCCC)\cite{penrose1969gravitational} and the Strong Cosmic Censorship Conjecture (SCCC)\cite{penrose1979singularities}. First, the WCCC asserts that spacetime singularities should be hidden behind the event horizon of a black hole, preventing external observers from directly observing the singularity. This implies that outside the black hole, the predictive power of general relativity remains reliable, as the event horizon effectively limits the influence of the singularity. Theoretical research on this conjecture has been tested in various black hole spacetime models. For example, in both static and rotating black holes, the WCCC has been supported\cite{liang2019weak}\cite{zhao2024weak}\cite{meng2023test}\cite{Zhao:2024qzg}\cite{Zhao:2023vxq}, theoretically ruling out the existence of naked singularities. However, in charged or more complex rotating black holes, naked singularities may still exist\cite{richartz2011challenging}\cite{gao2022testing}\cite{hod2013cosmic}, indicating that further research and verification are needed for the WCCC under these more complex conditions.

In contrast, the SCCC advocates that physical processes described by general relativity should remain predictable throughout spacetime. Specifically, SCCC posits that the Cauchy horizon is unstable, and any initial perturbations near the Cauchy horizon will be infinitely amplified due to the blue-shift effect, causing the stress-energy tensor to diverge, thereby disrupting the stability of the Cauchy horizon. As a result of this divergence, the Cauchy horizon cannot serve as a well-defined causal boundary, rendering it a non-extendable spacetime structure. This prevents internal black hole events from influencing external regions, ensuring that the causal structure of general relativity remains intact and predictable near the Cauchy horizon and outside the black hole, thus preserving the physical continuity and determinism of the external region. The validity of SCCC varies across different spacetime backgrounds, with some cases supporting the conjecture.For instance, in asymptotically flat Reissner-Nordström and Kerr black holes, perturbations experience an exponential blueshift effect as they approach the Cauchy horizon, causing the stress-energy tensor to diverge, thereby preventing the spacetime metric from extending beyond the Cauchy horizon and ensuring the determinism and causality of physical laws both inside and outside the black hole\cite{gurriaran2024precise}. Moreover, although perturbation modes may partially decay in rotating Kerr-de Sitter black holes, the blueshift effect is generally sufficient to induce instability at the Cauchy horizon, thus supporting the SCCC\cite{burko2016cauchy}\cite{markovic1995classical}\cite{sbierski2023instability}.

However, under certain extreme conditions, the SCCC may be violated. For example, in nearly extremal charged Reissner-Nordström-de Sitter (RNdS) black holes, the decay effect of perturbations may be stronger than the gravitational blueshift effect, thereby violating the SCCC\cite{mo2018strong}\cite{ben2018efficient}\cite{dias2018strong}\cite{jiang2023restoring}. Similar situations may also occur in other near-extremal black holes, such as near-extremal Kerr-Newman and Kerr-Newman-de Sitter black holes\cite{hod2018quasinormal}\cite{davey2024strong}\cite{zhang2021strong}\cite{gwak2018thermodynamics}. In these cases, the decay effect of perturbations is stronger than the gravitational blueshift effect. the Cauchy horizon may not exhibit sufficient divergence, allowing the spacetime metric to extend beyond the Cauchy horizon, ultimately leading to the failure of the SCCC.

In this article, we will briefly introduce a special type of black hole solution short-hair black holes\cite{david1997black}. The uniqueness of these black holes lies in their extremely short characteristic parameters, which are confined near the event horizon of the black hole and are difficult to detect through long-distance measurements.
This phenomenon challenges the traditional "no-hair theorem"\cite{israel1967event}\cite{carter1971axisymmetric}, which states that the properties of a black hole should depend only on its mass, angular momentum, and charge, with no localized characteristic parameters. However, through in-depth analysis, I find that short-hair black holes comply with the physical laws of general relativity and can exist stably under certain conditions. Additionally, the existence of short-hair black holes poses new challenges to the SCCC, as the internal structure of short-hair black holes is more complex than that of traditional black holes. This discovery prompts us to reassess the applicability of the SCCC, especially under extreme conditions, and encourages further research into the internal physical properties of black holes. Such research has the potential to deepen our understanding of the fundamental laws of the universe and advance general relativity and black hole physics.

This paper will test whether the short-hair black hole satisfies the SCCC through the perturbations of a neutral massless scalar field and a charged massive scalar field. In Section 2, we will introduce the research methods related to SCCC. In Section 3, we will briefly introduce the metric equations of the short-hair black hole and present the perturbation equations for the neutral massless scalar field and the charged massive scalar field, while also explaining the conditions under which the SCCC holds. In Section 4, we will present in detail the results of the WGC method and the WKB numerical method. Through an in-depth analysis of these results, we will determine under what conditions the SCCC is upheld or violated. The final section will summarize the entire paper, review the key findings of the research, and discuss the implications of these results for the SCCC, as well as expectations for future research on the SCCC.

\section{\label{sec:level2}NUMERICAL METHODS}

Currently, there are multiple methods to test the SCCC, but this article will mainly focus on the Weak Gravity Conjecture (WGC)\cite{arkani2007string} and the WKB approximation method\cite{wentzel1926verallgemeinerung}\cite{kramers1926wellenmechanik}\cite{walsh1926photometry} to test the SCCC in the context of hairy black holes. We examine SCCC by analyzing the scalar field perturbations of the black hole. The WKB method analyzes the asymptotic behavior of black hole perturbation modes, helping us understand the evolution of scalar field perturbations in the context of hairy black holes, thereby testing the validity of SCCC\cite{cardoso2018j}. The WGC method, by studying perturbation behavior under weak gravitational coupling conditions, explores the stability of hairy black holes and provides critical theoretical support for verifying SCCC\cite{crisford2018testing}\cite{sadeghi2023strong}. The combination of these two methods integrates mathematical derivations and physical model analysis. Their application in the context of hairy black holes plays an important role in understanding whether SCCC holds under extreme physical conditions.

\subsection{\label{sec:level2.1} WKB method}

Quasi-normal modes (QNMs) research holds significant importance in astrophysics because these oscillation modes, which are produced when black holes are perturbed, can reveal key physical properties of black holes. The QNMs spectrum consists of complex frequencies, where the real part represents the oscillation’s natural frequency, and the imaginary part determines the rate of decay of the oscillation. By analyzing these frequencies, one can infer the relevant physical properties of the black hole, thereby verifying the correctness of general relativity's predictions. Moreover, different types of black holes exhibit distinct QNMs spectral features\cite{liu2021ringing}.These features allow observations of QNMs frequencies not only to probe the physical properties of black holes but also to distinguish between different types of black holes.Furthermore, QNMs research plays an important role in testing the SCCC. By analyzing the decay characteristics of black hole perturbations, the stability of the Cauchy horizon can be evaluated, thereby indirectly testing the SCCC and further supporting the theoretical framework of general relativity.

In order to obtain the QNMs frequencies of the short-hair black hole under scalar field perturbations more accurately, we will adopt the WKB method to calculate the QNMs frequencies of the black hole, which is an effective asymptotic analysis method and widely used in the study of quantum mechanics and wave equations. The WKB method approximates the solution of complex differential equations by expressing the solution as a rapidly oscillating function, using asymptotic expansions. It is particularly suited for systems with high frequencies or short wavelengths. In the strong gravitational fields near black holes, wave equations often have complex potential forms, and the WKB method can effectively handle these situations. Particularly in analyzing QNMs frequencies, the WKB method can accommodate multiple boundary conditions and provide precise results in the study of high-frequency modes. By extending to higher-order terms\cite{iyer1987black}\cite{konoplya2003quasinormal},the WKB method can further improve the accuracy of the solution, helping researchers gain a deeper understanding of the response mechanisms of black holes under external perturbations, assess the stability of black holes, and evaluate their role in cosmic evolution. This is of great significance for verifying general relativity and exploring the physical properties of black holes.

Furthermore, the WKB method plays an important role in the theoretical analysis of black hole gravitational wave signals. The waveform data captured by gravitational wave detectors, such as LIGO and Virgo\cite{cardoso2018j}\cite{singh2024greybody}, contains the QNMs frequencies of black holes. By comparing the observed QNMs frequencies with those calculated theoretically, scientists can verify the correctness of general relativity’s predictions and explore new physical phenomena. The WKB method provides an approximate analytical tool for studying the oscillation modes of black holes by calculating these QNM frequencies, effectively describing the propagation and decay of perturbations in the black hole background. This, in turn, reveals the stability of black holes and their evolution under various conditions. According to the Schrödinger-like equation, we obtain the QNMs frequencies using the WKB method.

\begin{equation}
iK - \left(n + \frac{1}{2}\right)  - \Lambda(n) = \Omega(n)
\label{1}
\end{equation}
which
\begin{equation}
K = \frac{V_0}{\sqrt{2V_0^{(2)}}}
\label{2}
\end{equation}

\begin{equation}
\Lambda(n) = \frac{1}{\sqrt{2V_0^{(2)}}} \left[ \frac{(a^2 + \frac{1}{4})V_0^{(4)}}{8 V_0^{(2)} }- \frac{(60a^2 + 7)}{288} \left( \frac{V_0^{(3)}}{V_0^{(2)}} \right)^2 \right]
\label{3}
\end{equation}

\begin{widetext} 
	\begin{align}
		\Omega(n) = & \, \frac{n + \frac{1}{2}}{2V_0^{(2)}} \left[ \frac{5(188a^2 + 77)}{6912} \left( \frac{V_0^{(3)}}{V_0^{(2)}} \right)^4 - \frac{(100a^2 + 51)}{384} \frac{(V_0^{(3)})^2V_0^{(4)}}{(V_0^{(2)})^3} \right] \nonumber \\
		& + \frac{n + \frac{1}{2}}{2V_0^{(2)}} \left[ \frac{(68a^2 + 67)}{2304} \left( \frac{V_0^{(4)}}{V_0^{(2)}} \right)^2 + \frac{(28a^2 + 19)}{288} \frac{V_0^{(3)}V_0^{(5)}}{(V_0^{(2)})^2} - \frac{(4a^2 + 5)}{288} \frac{V_0^{(6)}}{V_0^{(2)}} \right]
	\end{align}
\end{widetext}

In the above equation, \( V_0^{(k)} = \left( \frac{d^k V}{dx^k} \right) \bigg|_{r = r_0} \) represents the derivative of the effective potential of the equation, and \( a = n + \frac{1}{2} \). Therefore, from equation (1), we can derive:
\begin{align}
	\omega^2 ={} & \left[ V_0 + \left( -2V_0^{(2)} \right)^{1/2} \tilde{\Lambda}(n) \right] \notag \\
	& - i \left( n + \frac{1}{2} \right) \left( -2V_0^{(2)} \right)^{1/2} \left[ 1 + \tilde{\Omega}(n) \right]
\end{align}
which
\begin{equation}
\tilde{\Lambda} = \frac{\Lambda}{i}, \quad \tilde{\Omega} = \frac{\Omega}{\left( n + \frac{1}{2} \right)}
\label{6}
\end{equation}
In equation (5), \( \omega \) corresponds to the result of the first-order WKB approximation. Equation (1) applies to any physical problem governed by the Schrödinger-like equation:
\begin{equation}
\frac{d^2 \psi}{dx^2} + V(x) \psi(x) = 0
\label{7}
\end{equation}
and the proper boundary conditions for the positive mode. In particular, it is applicable to the determination of quantum mechanical resonances near the top of a one-dimensional potential barrier.

The WKB method is an effective asymptotic analysis tool, widely used for handling complex differential equations, particularly excelling in high-frequency and short-wavelength systems. This method is especially important in analyzing the QNMs frequencies of black holes. By analyzing these frequencies, one can gain a deeper understanding of the response mechanism of black holes under perturbations. This not only helps in verifying the predictions of general relativity but also plays a crucial role in testing the SCCC. By calculating the QNMs frequencies using the WKB method, researchers can assess the stability of the Cauchy horizon, indirectly supporting the correctness of general relativity. In the future, with further development of the WKB method, it will play a greater role in a broader range of physical systems, reinforcing our understanding of black hole behavior in the universe.

\subsection{\label{sec:level2.2}Weak Gravity Conjecture}

Since 2007, the study of the  Weak Gravity Conjecture(WGC) has attracted widespread attention, particularly after 2010, when research interest in it significantly increased. The WGC is a hypothesis regarding quantum gravitational interactions, which can be simply described as all standard interaction forces must be stronger than gravity, expressed as:
\begin{equation}
F_{\text{gravity}} \leq F_{\text{any}}
\label{8}
\end{equation}
This conjecture aims to provide experimentally verifiable predictions by constraining certain cosmological and particle physics models, and it is tested through quantum gravity effects in the low-energy regime. Furthermore, the WGC requires that there must exist an object in any gauge theory that satisfies the following conditions:
\begin{equation}
\frac{|q|}{m} \geq \frac{|Q|}{M} \bigg|_{\text{ext}}
\label{9}
\end{equation}
In this context, \( \left. \frac{|Q|}{M} \right|_{\text{ext}} \) represents the charge-to-mass ratio of the extremal black hole under study. One of the core ideas proposed by the WGC is that in a theory of quantum gravity, there must exist a particle whose charge-to-mass ratio is higher than that of an extremal black hole, which implies that the charge of this particle must be greater than its mass. However, for massless scalar fields, the WGC cannot be directly applied.

According to the WGC, a generally extremal charged black hole may undergo some form of decay while maintaining stability. This decay could produce new black holes and other particles. 

To avoid the formation of a naked singularity, the WGC suggests that these decay products, specifically stable black holes, must satisfy the condition \( M > Q \), meaning the mass must be greater than the charge. This aligns with the requirements of the WCCC, which stipulates that stable black holes cannot form naked singularities and must be enclosed by an event horizon. However, some decay products may exhibit characteristics where the charge exceeds the mass, i.e., \( Q > M \).

To avoid the formation of a naked singularity, a stable and conventional black hole must satisfy the condition that its mass is greater than its charge, i.e., \( M > Q \). This is consistent with the requirement of the WCCC, which states that a stable black hole cannot form a naked singularity and that the singularity must be enclosed by the event horizon. However, certain decay products of extremely charged black holes may exhibit characteristics where the charge exceeds the mass, i.e., \( Q > M \). Such products cannot form stable black holes, as they would violate the WCCC. According to the WGC, these products should be particles rather than black holes to ensure consistency with physical laws.

The WGC has made significant progress in various research fields, but many unresolved questions remain. The WGC is not a unified theory but rather consists of multiple related conjectures, with the core idea being that the interaction forces must be stronger than gravity. Currently, the Tower Weak Gravity Conjecture\cite{ibanez2017constraining}\cite{benakli2020u} and the Sublattice Weak Gravity Conjecture\cite{caceres2019constraining}\cite{harlow2023weak} are the more deeply studied versions. Although some research has proposed methods to prove the WGC, these methods either lack precise predictions with \( O(1) \) factors or rely on unverified assumptions\cite{hamada2019weak}\cite{bellazzini2019positivity}. As a result, the different versions of the WGC have yet to form a unified expression, making its applications and influence in other fields difficult to determine. However, the WGC may have potential in testing the SCCC, indirectly supporting general relativity by verifying the stability of black holes. Further research could bring new breakthroughs in fields such as cosmology and particle physics.

\section{\label{sec:level3} Scalar Field Perturbation and Strong Cosmic Censorship Conjecture}
Black holes are significant celestial objects in the study of general relativity. They not only provide an ideal platform for testing the validity of general relativity but also reveal the behavior of gravity under extreme conditions. By studying the perturbations of black holes by scalar fields, we can delve deeper into the SCCC. This conjecture posits that, with few exceptions, the spacetime structure of general relativity should remain predictable. Verifying the SCCC not only helps confirm the applicability of general relativity under extreme conditions but also has profound implications for understanding the fundamental nature of gravity and the evolution of the universe.
\subsection{\label{sec:level3.1}Short-Hair black hole}
A short-hair black hole\cite{david1997black} is a special black hole solution, which demonstrates that under certain conditions, black holes can possess "hair" properties, thereby expanding our understanding of black hole structure. The traditional no-hair theorem suggests that black holes are described only by three external parameters: mass, charge, and angular momentum. However, research on hairy black holes indicates that black holes can also be described by other parameters. Although these parameters cannot be detected from afar, they play a crucial role near the event horizon. Studying short-hair black holes not only helps us explore the behavior of black holes under extreme conditions but also provides important theoretical support for verifying the SCCC.

For a static spherically symmetric black hole, its metric can be written as:
\begin{equation}
ds^2 = -f(r) dt^2 + \frac{1}{g(r)} dr^2 + r^2 (d\theta^2 + \sin^2\theta \, d\phi^2)
\label{10}
\end{equation}
By calculating the corresponding components of the Einstein tensor, we obtain \( G_t^t = G_r^r = \frac{r f' - 1 + f}{r^2} \) and \( G_\theta^\theta = G_\phi^\phi = \frac{r f'' + 2 f'}{2r} \). Next, we introduce an anisotropic fluid and set its equation of state as \( \rho(n) = C n^{k+1} \), and the pressure is obtained as \( P = n \frac{\partial \rho}{\partial n} - \rho = k \rho \). Then, substituting the fluid's equation of state into the components of the Einstein equation \( r f' - 1 + f = -8 \pi r^2 \rho \) and \( r f'' + 2 f' = 16 \pi r P \), we get the expressions for the energy density and pressure \( \rho = \frac{Q^{2k} (2k - 1)}{8 \pi r^{2k + 2}} \) and \( P = k \rho \). Finally, we can obtain the metric function for the hairy black hole.
\begin{equation}
f(r) = g(r) = 1 - \frac{2M}{r} + \frac{Q^{2k}}{r^{2k}}
\label{11}
\end{equation}
In classical general relativity, the metric equation of the Reissner-Nordström black hole is given by:
\(f(r) = 1 - \frac{2M}{r} + \frac{Q^2}{r^2}\)where \( M \) is the mass of the black hole and \( Q \) is the charge. To ensure the existence of an event horizon and avoid the appearance of a naked singularity, the condition \( M \geq Q \) must be satisfied. This model is relatively simple and does not take into account complex factors such as higher dimensions or modified gravity. The metric equation (11) we derived extends this model, making it applicable to more complex black hole models, such as those in higher-dimensional spaces or modified gravity theories.

The event horizon radius of a hairy black hole is given by \( g^{rr} = 0 \), i.e., \( f(r) = 0 \). If equation (11) has degenerate real roots, then the hairy black hole is in an extremal state. This not only requires the metric function \( f(r) = 0 \), but also the condition that the derivative \( \left. \frac{df}{dr} \right|_{r = r_h} = 0 \) must be satisfied at this point. From \( f(r) = 0 \) and \( \left. \frac{df}{dr} \right|_{r = r_h} = 0 \), we can obtain the final relationship between the black hole mass \( M \) and the black hole charge \( Q \).
\begin{equation}
M \geq k \cdot (2k - 1)^{\frac{1 - 2k}{2k}} \cdot Q
\label{12}
\end{equation}
When equation (12) takes the equality sign, the charge of the hairy black hole reaches its maximum value, denoted as \( Q_{\text{max}} \).
\begin{figure*}
	\centering
	\resizebox{\linewidth}{!}{\includegraphics*{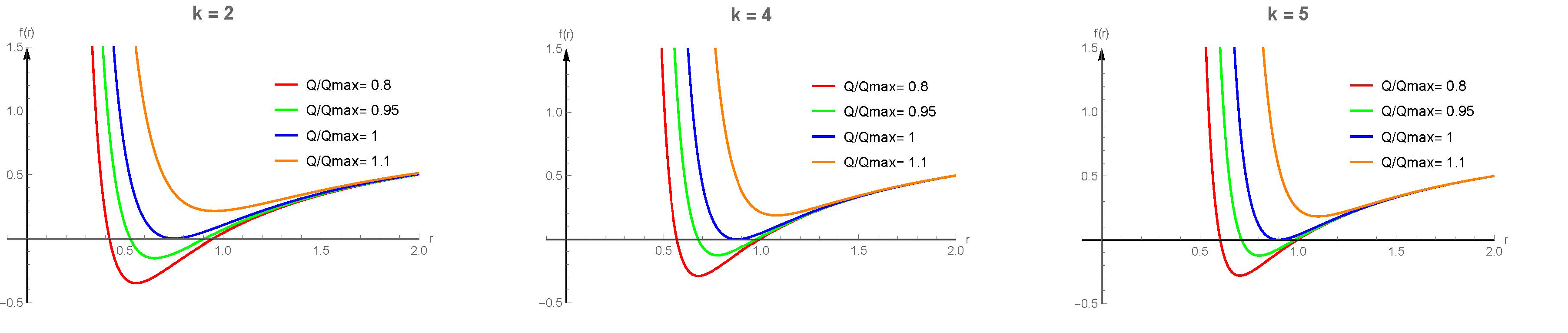}}
	\captionsetup{justification=raggedright, singlelinecheck=false}  
	\caption{Shows the existence of the event horizon under different values of the order $k$ and varying ratios of $Q/Q_{\text{max}}$ for the metric function.}
	\label{fig:7}
\end{figure*}

In FIG 1, we show that when the black hole's charge \( Q \) is less than its maximum value \( Q_{\text{max}} \), the black hole has two horizons for different values of the order \( k \). These are the event horizon \( r_{h_+} \) and the Cauchy horizon \( r_{h_-} \). The event horizon is a boundary surrounding a black hole, beyond which no matter or information can return or escape from the gravitational pull of the black hole. This means that the external world cannot access any information from inside the event horizon. The Cauchy horizon, located within the event horizon, is another critical boundary inside the black hole, marking the limit of the spacetime structure. Beyond the Cauchy horizon, existing physical laws (such as general relativity) may no longer apply, and the predictability of causal relationships breaks down, rendering the spacetime structure uncalculable and incomprehensible. Consequently, time and space may become highly irregular, and physical processes near the singularity will remain unknown.

We can also observe from FIG 1 that when the charge \( Q \) of the hairy black hole reaches its maximum value \( Q_{\text{max}} \), the black hole only has one horizon, even for different values of the order \( k \). This indicates that the black hole is in an extremal state. When the charge \( Q \) of the hairy black hole exceeds its maximum value \( Q_{\text{max}} \), the event horizon of the black hole will disappear, exposing the singularity inside. This situation violates the cosmic censorship conjecture, suggesting that general relativity may fail to provide a complete description under extreme conditions, such as a naked singularity. This indicates that under such circumstances, general relativity needs to be combined with other theories to fully explain these physical phenomena.

\subsection{\label{sec:level3.2}Neutral massless scalar field}
In this section, we study the evolution of perturbations of a neutral massless scalar field in the spacetime background of a hairy black hole. The evolution of the scalar field in curved spacetime is governed by the Klein-Gordon equation:
\begin{equation}
\frac{1}{\sqrt{-g}} \partial_\mu \left( g^{\mu \nu} \sqrt{-g} \, \partial_\nu \Phi \right) = 0
\label{13}
\end{equation}
Due to the spherical symmetry of the spacetime, we can express the scalar field using spherical harmonics.
\begin{equation}
\Phi(t, r, \theta, \phi) = \sum_{m \ell} \frac{\varphi(r)}{r} Y_{\ell m}(\theta, \phi) e^{-i \omega t}
\label{14}
\end{equation}
Here, \( \omega \) is the frequency of \( \Phi \), and \( \phi(r) \) is the radial wave function. \( \ell \) is the angular momentum quantum number, which only takes non-negative integers. \( m \) is the magnetic quantum number, and throughout this paper, we always set it to zero. \( Y_{\ell m}(\theta, \phi) \) is the spherical harmonic function. Substituting \( \Phi \) into the Klein-Gordon equation (13), we get:
\begin{widetext} 
	\begin{equation}\label{Omega_n}
		\begin{aligned}
			\left[ 
			-\frac{r^2}{f(r)} \frac{\partial^2}{\partial t^2} + \frac{\partial}{\partial r} \left( r^2 f(r) \frac{\partial}{\partial r} \right) + \frac{1}{\sin \theta} \frac{\partial}{\partial \theta} \left( \sin \theta \frac{\partial}{\partial \theta} \right) + \frac{1}{\sin^2 \theta} \frac{\partial^2}{\partial \phi^2}
			\right] \Phi = 0
		\end{aligned}
	\end{equation}
\end{widetext}
We can use the method of separation of variables to decompose the equation into a radial equation and an angular equation. The angular equation is:
\begin{equation}
	\begin{aligned}
		\left[ \frac{1}{\sin \theta} \frac{\partial}{\partial \theta} \left( \sin \theta \frac{\partial}{\partial \theta} \right) + \frac{1}{\sin^2 \theta} \frac{\partial^2}{\partial \phi^2} \right] Y_{\ell m}(\theta, \phi) \\
		= \ell(\ell+1) Y_{\ell m}(\theta, \phi)
	\end{aligned}
\end{equation}
The radial equation is:
\begin{equation}
	\begin{aligned}
		\left[ -\frac{r^2}{f(r)} \frac{\partial^2}{\partial t^2} + \frac{\partial}{\partial r} \left( r^2 f(r) \frac{\partial}{\partial r} \right) \right] \frac{\varphi(r)}{r} e^{-i \omega t} \\
		= -\ell(\ell+1) \frac{\varphi(r)}{r} e^{-i \omega t}
	\end{aligned}
\end{equation}
We are only interested in the radial equation, which can be simplified to:
\begin{equation}
\left( \frac{d^2}{dr_*^2} + \omega^2 - V_{\text{eff-1}}(r) \right) \varphi(r) = 0
\label{18}
\end{equation}
Here, \( dr_* \) is the tortoise coordinate, which is defined by the differential equation \( dr_* = \frac{dr}{f(r)} \). Since \( f(r) \) is usually a nonlinear function, the coordinates \( r \) and the tortoise coordinate \( r_* \) generally do not have a linear relationship. The range of \( r_* \) extends from \( -\infty \) at the horizon to \( +\infty \) at spatial infinity. \( V_{\text{eff-1}}(r) \) represents the potential, which is expressed as:
\begin{equation}
V_{\text{eff-1}}(r)= f(r) \left[ \frac{\ell(\ell+1)}{r^2} + \frac{f'(r)}{r} \right]
\label{19}
\end{equation}
Due to the physical properties of the black hole event horizon, any incoming wave reaching the event horizon will be completely absorbed, with no reflection. Therefore, the boundary condition at the event horizon is that of a purely incoming wave. At spatial infinity, the boundary condition is that of a purely outgoing wave.
\begin{equation}
\varphi(r) \sim
\begin{cases}
	e^{-i \omega r_*}, & r_* \to -\infty \\
	e^{i \omega r_*}, & r_* \to +\infty
\end{cases}
\label{20}
\end{equation}
For most cases, the effective potential gradually approaches zero at spatial infinity, causing the perturbation modes to decay and vanish at infinity. Based on this characteristic, we can solve the perturbation equation by imposing appropriate boundary conditions. Ultimately, we obtain a discrete set of QNMs frequencies, which describe the characteristic oscillation modes of the system.
\subsection{\label{sec:level3.3}Charged massive scalar field}
The Klein-Gordon equation for a charged massive scalar field can be written as:
\begin{equation}
	\begin{aligned}
		\frac{1}{\sqrt{-g}} \partial_\mu \left( g^{\mu \nu} \sqrt{-g} \partial_\nu \Phi \right) 
		- 2 i q g^{\mu \nu} A_\mu \partial_\nu \Phi \\
		- q^2 g^{\mu \nu} A_\mu A_\nu \Phi - m^2 \Phi = 0
	\end{aligned}
\end{equation}
Here, \( m \) and \( q \) are the mass and charge of the scalar field, respectively. The electromagnetic four-potential in four-dimensional spacetime is given by \( A_\mu = \left( -\frac{Q}{r}, 0, 0, 0 \right) \), which describes the electric potential generated by the black hole at the radial distance \( r \), where \( Q \) is the charge of the black hole.We can derive the radial equation of the scalar field from the relationship (21) and the defined scalar field \( \Phi \):
\begin{equation}
\frac{d^2 \varphi(r)}{dr_*^2} + V_{\text{eff-2}}(r) \varphi(r) = 0
\label{22}
\end{equation}
The relevant potential function equation becomes:
\begin{equation}
	\begin{aligned}
		V_{\text{eff-2}}(r) = \omega^2 - f(r) \left[ \frac{\ell(\ell+1)}{r^2} + \frac{f'(r)}{r} \right] \\
		+ \frac{q^2 Q^2}{r^2} - \frac{2 \omega q Q}{r} + m^2 f(r)
	\end{aligned}
\end{equation}

Due to the physical properties of the black hole event horizon, any incoming wave reaching the event horizon will be completely absorbed, with no reflection. Therefore, the boundary condition at the event horizon is that of a purely incoming wave. In addition, the boundary condition at spatial infinity is that of a purely outgoing wave.
\begin{equation}
\varphi(r) \sim 
\begin{cases}
	e^{-i \left( \omega - \frac{qQ}{r_{h_+}} \right) r_*}, & r_* \to -\infty \\
	e^{i \omega r_*}, & r_* \to +\infty
\end{cases}
\label{24}
\end{equation}
The discrete spectrum of frequencies can be obtained from boundary conditions. This discrete spectrum is usually referred to as the QNMs frequencies, which describe the oscillatory behavior of a system under specific conditions. The imaginary part of the QNMs frequencies is related to the rate of decay or growth of the oscillatory modes, while the real part represents the fundamental oscillation frequency of the system, thus comprehensively reflecting the resonance characteristics of the system.
\subsection{\label{sec:level3.4}Strong Cosmic Censorship Conjecture}
In mathematical terms, the decay expression for black hole perturbation modes is given by \(\varphi \sim \exp(-\omega u)\varphi_0\), where \(\omega\) represents the imaginary part of the QNMs frequencies, indicating the rate of decay of black hole perturbations. The larger the imaginary part of the QNMs frequency, the faster the decay . Here, \(u\) is a parameter describing the time evolution, expressed as \(u = t + r_*\), where \(t\) is the time coordinate, and \(r_*\) is the so-called "tortoise coordinate," which is related to the radial coordinate \(r\) through an integral and is used to handle spacetime characteristics near the horizon. Additionally, the expression for the blueshift effect of the scalar field near the horizon is given by \(|\varphi_{r_h}|^2 \sim \exp(k_i u)|\varphi_0|^2\). In this context, \(k_i\) represents the surface gravity of the horizon, defined as the gravitational acceleration at the horizon, reflecting the strength of the gravitational field near the horizon. The horizon here can be a Cauchy horizon, event horizon, etc., depending on the specific case considered. The specific expression for the surface gravity \(k_i\) is:
\begin{equation}
k_i = \left| \frac{1}{2} f'(r_i) \right|
\label{25}
\end{equation}

The larger the surface gravity of the black hole, the stronger the gravitational field at the horizon. For example, perturbations near the Cauchy horizon may be amplified due to the dynamical instability of the inner horizon, which could lead to a divergence in the perturbation modes, closely related to the validity of the SCCC. The SCCC asserts that under general physical conditions, solutions to Einstein's field equations are deterministic, meaning the evolution of the future can be uniquely determined from initial conditions. The existence of the Cauchy horizon may lead to the non-uniqueness of the solution. If the perturbation modes diverge at the Cauchy horizon, it means that the solution may not be extendable beyond the Cauchy horizon, thereby leading to a breakdown of the deterministic nature of spacetime, supporting the SCCC. Therefore, the parameters \(k_i\) and \(\omega\) play a crucial role in verifying the stability of physical theories inside black holes and the validity of the SCCC.

QNMs describe specific modes of decay in black hole or spacetime perturbations that satisfy particular boundary conditions. For hairy black holes, their characteristic frequencies \( \omega \) are discrete. The extendibility of the solution depends on the condition of local square integrability, which means that the physical acceptability of the perturbation solutions relies on their smoothness and boundedness. In other words, if these solutions satisfy the square integrability condition in a local region, i.e., the square of the derivatives is integrable over a finite region, then these solutions can be extended beyond the Cauchy horizon, thereby remaining valid over a larger spacetime domain. This integrability condition ensures that the behavior of the solutions near the Cauchy horizon is controlled, which in turn supports the validity of the SCCC.

The validity of the SCCC is related to the ratio \( \beta \):
\begin{equation}
\beta = -\frac{\text{Im}\, \omega}{k_i}
\label{26}
\end{equation}
When the ratio of the imaginary part of the perturbation mode to the surface gravity of the horizon \(\text{Im} \, \omega / \kappa_i\) is less than \(1/2\), it implies that the perturbation modes will diverge, ensuring the validity of the SCCC. For instance, in the case of a black hole's Cauchy horizon, when the ratio \(\beta\) is smaller than the critical value \((1/2)\), the perturbations inside the black hole cannot remain finite, leading to the divergence of physical quantities at the Cauchy horizon. This prevents the solution from being mathematically extended beyond the Cauchy horizon, thus preserving the determinism of spacetime. However, when \(\text{Im} \, \omega / \kappa_i\) is greater than \(1/2\), the perturbation modes will not diverge, meaning the perturbations can be extended beyond the Cauchy horizon, thereby violating the SCCC.

\section{\label{sec:level4}Results Analysis}
This chapter provides a detailed analysis of the effects of neutral massless scalar fields and charged massive scalar fields on short-hair black holes. By thoroughly investigating the evolution behavior of these two types of scalar fields around black holes, we explore the conditions under which the SCCC can be maintained or possibly violated\cite{hod2020proof}\cite{ahmed2022weak}. In this study, we employ both the WKB method and the WGC, which are widely applied in black hole physics research. The WKB method not only helps to obtain accurate numerical results, but it is also crucial in exploring quantum effects, while the WGC provides theoretical support for understanding the relevant physical mechanisms. By combining these two methods, we determine the necessary conditions for short-hair black holes to obey the SCCC under specific scenarios. Finally, through graphical representations, we present these results to better understand the effects of scalar fields on short-hair black holes and the nature of the related physical phenomena.

\subsection{\label{sec:level4.1}Results for neutral massless scalar fields}
In this section, we will use a neutral massless scalar field to verify the short-hair black hole and explore whether it complies with the SCCC under different conditions. First, we set the black hole’s mass to $M=0.5$. This choice simplifies subsequent calculations and ensures that the parameter selection is physically reasonable and representative. Next, we need to determine the order $k$ of the metric equation for the short-hair black hole and, based on the above conditions, calculate the maximum charge of the black hole, denoted as $Q_{\text{max}}$. We then use the WKB method\cite{konoplya2019higher} to systematically vary the ratio $Q/Q_{\text{max}}$ and the angular momentum quantum number $l$ to compute the QNMs frequencies of the black hole under these conditions. Finally, we compare the imaginary part of the QNMs frequency $\omega$ with the surface gravity of the Cauchy horizon $k_{r_{h_-}}$ to determine whether the SCCC is satisfied.

It is particularly important to emphasize that we are using the surface gravity of the Cauchy horizon $k_{r_{h_-}}$ here, rather than the surface gravity of the event horizon $k_{r_{h_+}}$, because the SCCC primarily concerns the stability of the Cauchy horizon.If the Cauchy horizon is unstable, this could lead to spacetime indeterminacy, making certain physical processes unpredictable and potentially challenging the classical understanding of causality in general relativity. Therefore, by comparing the imaginary part of the black hole’s QNMs frequency with the surface gravity of the Cauchy horizon, we can effectively assess the stability of the black hole’s interior and verify the validity of the SCCC. The WKB method, when used to study the SCCC, focuses more on the surface gravity at the Cauchy horizon because it directly affects the causal structure and stability of the black hole’s interior.

The table below shows the results of the ratio $\beta$ calculated using the WKB method under different conditions of the charge ratio $Q/Q_{\text{max}}$ and the angular momentum quantum number $l$. From these data, we can intuitively observe whether the SCCC is violated under specific conditions and which parameter combinations have the greatest impact on the SCCC.

\begin{table}[htbp] 
	\centering
	\begin{minipage}{0.5\textwidth}
		\centering
		\begin{tabular}{|c|c|c|c|c|}
			\hline
			$Q/Q_{\text{max}}$ & $l=0$ & $l=1$ & $l=10$ & $l=20$ \\
			\hline
			0.7   & 0.037794 & 0.028997 & 0.026027 & 0.025977 \\
			\hline
			0.8   & 0.058388 & 0.050777 & 0.050247 & 0.0524 \\
			\hline
			0.9   & 0.129969 & 0.113602 & 0.112995 & 0.112987 \\
			\hline
			0.95  & 0.242129 & 0.209021 & 0.208394 & 0.208388 \\
			\hline
			0.99  & 0.751882 & 0.63338  & 0.631158 & 0.631139 \\
			\hline
		\end{tabular}
			\captionsetup{justification=raggedright, singlelinecheck=false}
		\caption{Calculated values of $\beta$ for different angular momentum quantum numbers $l$ and charge ratio $Q/Q_{\text{max}}$, with the short-hair black hole metric equation for $k=2$ and $M=0.5$.}
		\label{tab:short_hair_black_hole1}
	\end{minipage}
	
	\vspace{1cm}
	
	\begin{minipage}{0.5\textwidth}
		\centering
			\begin{tabular}{|c|c|c|c|c|}
			\hline
			$Q/Q_{\text{max}}$ & $l=0$ & $l=1$ & $l=10$ & $l=20$ \\
			\hline
			0.7   & 0.032043 & 0.027381 & 0.026927 & 0.0269921 \\
			\hline
			0.8   & 0.059361 & 0.051026 & 0.05029  & 0.050281 \\
			\hline
			0.9   & 0.129652 & 0.111459 & 0.110292 & 0.110281 \\
			\hline
			0.95  & 0.314127 & 0.229354 & 0.202508 & 0.202073 \\
			\hline
			0.99  & 0.739963 & 0.615429 & 0.61007  & 0.610048 \\
			\hline
		\end{tabular}
		\captionsetup{justification=raggedright, singlelinecheck=false}
		\caption{Calculated values of $\beta$ for different angular momentum quantum numbers $l$ and charge ratio $Q/Q_{\text{max}}$, with the short-hair black hole metric equation for $k=4$ and $M=0.5$.}
		\label{tab:short_hair_black_hole2}
	\end{minipage}
	
	\vspace{1cm}
	
	\begin{minipage}{0.5\textwidth}
		\centering
		\begin{tabular}{|c|c|c|c|c|}
			\hline
			$Q/Q_{\text{max}}$ & $l=0$ & $l=1$ & $l=10$ & $l=20$ \\
			\hline
			0.7   & 0.028676 & 0.025502 & 0.025201 & 0.025194 \\
			\hline
			0.8   & 0.028398 & 0.025485 & 0.025165  & 0.025160 \\
			\hline
			0.9   & 0.115463 & 0.104538 & 0.102950 & 0.102931 \\
			\hline
			0.95  & 0.212343 & 0.192744 & 0.189148 & 0.189116 \\
			\hline
			0.99  & 0.638443 & 0.589299 & 0.575314  & 0.575206 \\
			\hline
		\end{tabular}
		\captionsetup{justification=raggedright, singlelinecheck=false}
		\caption{Calculated values of $\beta$ for different angular momentum quantum numbers $l$ and charge ratio $Q/Q_{\text{max}}$, with the short-hair black hole metric equation for $k=5$ and $M=0.5$.}
		\label{tab:short_hair_black_hole3}
	\end{minipage}
	
\end{table}

From TABLE \text{I}, \text{II} and \text{III}, we can observe that, with the order $k$ and the black hole mass $M$ fixed, as the charge ratio $Q/Q_{\text{max}}$ increases and the angular momentum quantum number $l$ decreases, the ratio $\beta$ becomes progressively larger and eventually exceeds $1/2$. This result implies that the physical process may deviate from classical expectations, leading to the failure of the SCCC. This finding challenges the fundamental assumptions in classical black hole theory and prompts us to reconsider the internal structure of black holes and the behavior of singularities. Next, we analyze the obtained results through graphical representation. We plot the variation of $\beta$ with the charge ratio $Q/Q_{\text{max}}$ after fixing the value of the order $k$ and the black hole mass $M$, as well as the variation of $\beta$ with $k$ after fixing the black hole mass $M$ and the charge ratio $Q/Q_{\text{max}}$.
\begin{figure*}
	\centering
	\resizebox{\linewidth}{!}{\includegraphics*{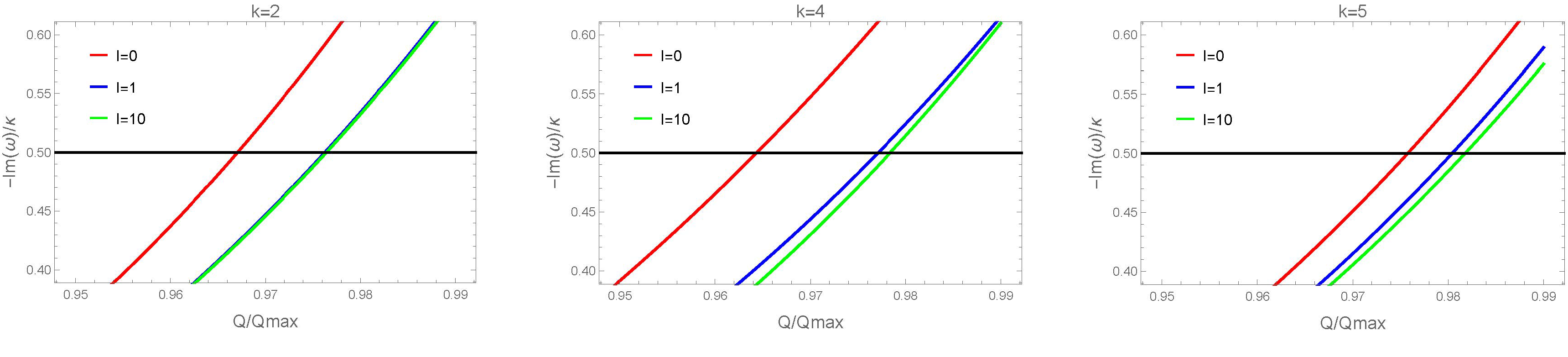}}
	\captionsetup{justification=raggedright, singlelinecheck=false}  
	\caption{The above image shows the variation of $\beta$ with $Q/Q_{\text{max}}$ under the condition that the black hole mass $M = 0.5$ and the value of the order $k$ are fixed.}
	\label{fig:7}
\end{figure*}

\begin{figure*}
	\centering
	\resizebox{\linewidth}{!}{\includegraphics*{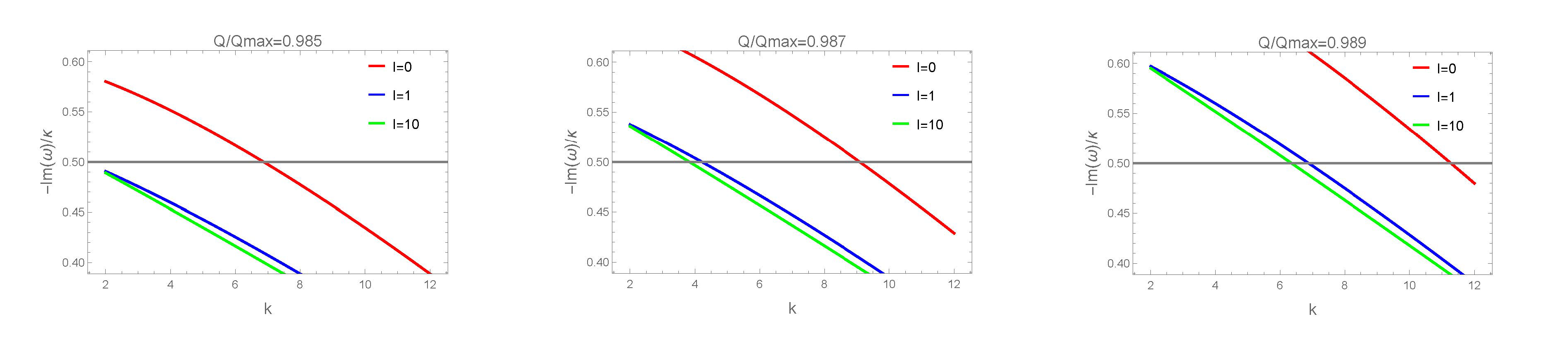}}
	\captionsetup{justification=raggedright, singlelinecheck=false}  
	\caption{Figure 3: The above image shows the variation of $\beta$ with $k$ under the condition that the black hole mass $M = 0.5$ and the ratio $Q/Q_{\text{max}}$ are fixed.}
	\label{fig:7}
\end{figure*}

To better explain this phenomenon, FIG 2 shows how the ratio $\beta$ varies with $Q/Q_{\text{max}}$ when the order $k$ and the black hole mass $M$ are fixed. We can observe that as the charge $Q$ gradually approaches its maximum value, the SCCC is always violated. This implies that in the context of short-hair black holes, the closer the charge $Q$ is to its maximum value, the more likely the SCCC is to be violated.In FIG 3, we plot the variation of the parameter $\beta$ with the order $k$ when the black hole mass $M$ and $Q/Q_{\text{max}}$ are fixed. It can be observed that although there are instances where the SCCC is violated, as the value of $k$ increases, the adherence to the SCCC becomes more likely. Therefore, the magnitude of the order \( k \) plays a crucial role in determining whether short-hair black holes comply with the SCCC, as it influences the evolution of scalar field perturbations near the Cauchy horizon. We also find that the larger the angular momentum quantum number $l$, the more likely the SCCC is to be upheld. This indicates that the angular momentum quantum number $l$ also plays an important role in maintaining the SCCC. In summary, when the black hole’s $Q/Q_{\text{max}}$ approaches its limit, the SCCC is always violated. However, the size of the order $k$ and the angular momentum quantum number $l$ can mitigate the likelihood of the SCCC being violated. These results demonstrate that whether the black hole charge $Q$ is close to its extreme value, and the size of the angular momentum quantum number $l$ and the order $k$, play key roles in exploring the physical properties of black holes and verifying the SCCC.

\subsection{\label{sec:level4.2}Results for charged massive scalar fields}
In this section, we will study the perturbation frequencies of charged massive scalar fields near short-hair black holes to verify the validity of the SCCC\cite{crisford2018testing}\cite{sadeghi2023strong}. To ensure the universality of the results, we adopt a dimensionless approach. Through this method, we eliminate the influence of specific parameters, making the results applicable to a broader range of physical scenarios.According to the fine-structure constant in physics, $e^2/c\hbar \simeq 1/137$, a slightly charged scalar field can satisfy the condition $qQ \gg 1$, where $q$ represents the charge of the scalar field and $Q$ represents the charge of the black hole. It is important to note that the electric field strength of a charged black hole is limited by the Schwinger effect\cite{schwinger1951gauge}. The Schwinger effect describes how, in extremely strong electric fields, virtual particle pairs in the vacuum can be excited into real particle pairs, leading to the discharge of the black hole.

To prevent this rapid discharge phenomenon, the electric field strength of the black hole must be below the critical electric field strength that would produce Schwinger discharge. Specifically, the electric field strength of the black hole should satisfy the condition $Q/r_{h_+} \ll m^2/q$, where $r_{h_+}$ is the radius of the black hole's event horizon, and $m$ and $q$ are the mass and charge of the scalar field, respectively\cite{carter1974charge}\cite{hod1998best}\cite{zaumen1974upper}. This inequality imposes an upper limit on the charge-to-mass ratio of the black hole to avoid the production of virtual particle pairs through the Schwinger effect, which would result in rapid black hole discharge.This condition ensures that, near the black hole, the behavior of the charged scalar field is primarily governed by its mass, with the effects of the charge being weaker, thereby laying the foundation for subsequent analysis. Based on this, we can define the following constraint conditions to simplify the analysis of QNMs of charged massive scalar field perturbations:
\begin{equation}
m^2 r_{h_+}^2 \gg l(l+1); \quad m^2 r_{h_+}^2 \gg 2k_{r_{h_+}} r_{h_+}
\label{27}
\end{equation}

The first condition ensures the dominance of the mass term in the effective potential, allowing us to reasonably neglect the contribution of angular momentum; the second condition guarantees the dominance of the mass term over the surface gravity term of the black hole, further simplifying the analysis. The surface gravity at the black hole’s event horizon, $k_{r_{h_+}}$, is a key parameter. In this region, we need to compute the imaginary part of the QNMs near the horizon of the short-hair black hole. We study the linear dynamics of charged massive particles near the short-hair black hole horizon using the radial potential (22).First, the electric potential in region (27) is treated as an effective potential, and then the imaginary part of the QNMs near the black hole horizon is solved using the WKB method. In this region, we assume that the point $r_0$ near the charged black hole’s event horizon has the maximum effective potential. Using equations (22), (23), and $V'(r_0) = 0$, the point with the maximum effective potential can be determined:
\begin{equation}
r_0 = \frac{q^2 Q^2}{qQ\omega - m^2 r_{h_+}^2 k_{r_{h_+}}}
\label{28}
\end{equation}
Using equations (2),(3), (4),(20), (27), and (28), we can obtain the following:
\begin{equation}
K \simeq \frac{k_{r_{h_+}}^2 m^4 r_{h_+}^4 qQ}{2 f_{r_0} \left( k_{r_{h_+}} m^2 r_{h_+}^2 - qQ \omega \right)^2}
\label{29}
\end{equation}
\begin{widetext} 
	\begin{equation}\label{Omega_n}
		\begin{aligned}
		\Lambda(n) \simeq \frac{k_{r_{h_+}}^2 m^4 \left[ 17 - 60 \left( n + \frac{1}{2} \right)^2 \right] r_{h_+}^4 + 2 k_{r_{h_+}} m \left[ 36 \left( n + \frac{1}{2} \right)^2 - 7 \right] q Q r_{h_+}^2 \omega f_{r_0}}{16 q Q \left( qQ \omega - 3 k_{r_{h_+}} m^2 r_{h_+}^2 \right)^2}
		\end{aligned}
	\end{equation}
\end{widetext}

\begin{widetext} 
	\begin{equation}\label{Omega_n}
		\begin{aligned}
			\Omega(n) \simeq \frac{15 k_{r_{h_+}}^4 m^8 \left[ 148 \left( n + \frac{1}{2} \right) - 41 \right] r_{h_+}^8 + 12 k_{r_{h_+}}^3 m^{6} \left[ 121 - 420 \left( n + \frac{1}{2} \right)^2 \right] q Q r_{h_+}^6 \omega  (- \left( n + \frac{1}{2} \right) Q^3 q^3 f_{r_0}^2)}{64 q^5 Q^5 \left( k_{r_{h_+}} m^2 r_{h_+}^2 - qQ \omega \right)^4}
		\end{aligned}
	\end{equation}
\end{widetext}

Next, we need to determine the value of the imaginary part of the QNMs in the system to further investigate the SCCC. For this, we can use a combination of equation (1) and equation (29),(30), (31)to calculate $\text{Im}(\omega)$.
\begin{equation}
	\begin{aligned}
		\omega \simeq \frac{qQ}{r_{h_+}} - \frac{2k_{r_{h_+}} m^2 r_{h_+}^2}{qQ} 
		\left[ 1 - \frac{14400}{11644} 
		\left( \frac{n + \frac{1}{2}}{qQ} f_{r_0} \right)^4 \right] \\
		- i \left[ 4 f_{r_0} k_{r_{h_+}} 
		\left( n + \frac{1}{2} \right) \frac{m^2 r_{h_+}^2}{q^2 Q^2} 
		\left( 1 - \frac{34qQ f_{r_0}^4}{11644} \right) \right]
	\end{aligned}
\end{equation}
When $r_0$ is very close to the event horizon ($r_{h_+}$), We can find that $f_{r_0} \ll 1$. By calculating the value of the imaginary part of the QNMs and its ratio with the surface gravity of the event horizon, we can verify whether the SCCC is upheld.
\begin{equation}
\beta = \frac{-\text{Im}(\omega)}{k_{r_{h_+}}} \simeq 2 f_{r_0} \frac{m^2 r_{h_+}^2}{q^2 Q^2} 
\left[ 1 - \frac{34 qQ f_{r_0}^4}{11664} \right]
\label{33}
\end{equation}

Since the event horizon is an important boundary of the black hole, choosing the surface gravity $k_{r_{h_+}}$ of the event horizon can more directly reflect the decay behavior of perturbations near the event horizon. By analyzing the ratio $\text{Im}(\omega)/k_{r_{h_+}})$, researchers can evaluate the behavior of perturbations near the event horizon and determine whether the SCCC is likely to hold.

In Reissner-Nordström-like black holes, the surface gravity $k_{r_{h_+}}
$ of the black hole’s event horizon is usually smaller than the surface gravity $k_{r_{h_-}}$ of its Cauchy horizon. This means that the ratio $\beta$ calculated using $k_{r_{h_+}}$ will be relatively large, thus providing a new perspective for testing the SCCC. If the ratio $\beta$ is still less than $1/2$ when using $k_{r_{h_+}}$, there is greater confidence that the SCCC holds, as this indicates that the perturbations would diverge as they approach the Cauchy horizon.

In summary, the reason for choosing the surface gravity $k_{r_{h_+}}
$ of the event horizon instead of the surface gravity $k_{r_{h_-}}
$ of the Cauchy horizon in equation (33) is because this allows the validity of the SCCC to be tested more effectively through the perturbations' behavior in the exterior region of the black hole. This choice ensures the reasonableness and consistency of the physical analysis, thereby more accurately reflecting the dynamical properties inside the black hole.

Since $f_{r_0} \ll 1$, in order to satisfy $\beta < 1/2$, We need to establish a new condition \(q^2 Q^2 > m^2 r_{h_+}^2\). Therefore, the following conditions are considered for the study of the SCCC:
\begin{equation}
\frac{q}{m} \geq \frac{r_{h_+}}{Q}
\label{34}
\end{equation}
From equation (34), it can be concluded that when $r_{h_+} \geq Q$, it will satisfy the WCCC. When $qQ < 2\sqrt{f_{r_0}  } m r_{h_+}$, the SCCC will be violated. Since $qQ \gg 1$ and $f_{r_0} \ll 1$, the mass of the scalar field and the radius of the black hole's event horizon must be sufficiently large.

Next, we study the metric equation (11). When $k_{r_{h_+}} = k_{r_{h_-}}
 = 0$, we can obtain the extremal values of the charge $Q$ and mass $M$ of the short-hair black hole.
\begin{equation}
Q_{\text{exe}} = \left( \frac{r_{h_+}^{2k}}{2k - 1} \right)^{\frac{1}{2k}} \cdot M_{\text{exe}} = \frac{k \cdot r_{h_+}}{2k - 1}
\label{35}
\end{equation}
Substituting equation (35) into equation (33) gives:
\begin{equation}
\beta \simeq 2 f_{r_0} \frac{m^2}{q^2} (2k - 1)^{\frac{1}{k}} 
\left[ 1 - \frac{34 q f_{r_0}^4}{11664} 
\left( \frac{r_{h_+}^{2k}}{2k - 1} \right)^{\frac{1}{2k}} \right]
\label{36}
\end{equation}
Based on the above relations, we can conclude that when $\frac{q^2}{m^2} > (2k - 1)^{\frac{1}{k}}$, the SCCC will be upheld. When $k > 1$, the WCCC will also be satisfied. In other words, when $k \geq 1$, according to the WGC, we will obtain the condition for short-hair black holes to satisfy the SCCC.
\begin{equation}
\frac{q}{m} \geq \left( 2k - 1 \right)^{\frac{1}{2k}}
\label{37}
\end{equation}
Therefore, according to the WGC, when $\frac{q}{m}$ satisfies equation (37), the SCCC will certainly be upheld. When $2\sqrt{f_{r_0}}(2k - 1)^{1/2k} > \frac{q}{m}$, the SCCC will certainly be violated.

\section{\label{sec:level5}Summary}
In this study, we investigate the applicability of the SCCC in the context of short-hair black holes, with a focus on analyzing the impact of this conjecture on the singularity problem in general relativity. According to the cosmic censorship conjecture, singularities should be concealed by the event horizon to ensure the predictability of general relativity, as the existence of naked singularities would lead to unpredictability in physics, thereby threatening the integrity of general relativity. Short-hair black holes, as a special class of black hole solutions, have the property that certain physical characteristics manifest only near the event horizon and become difficult to detect from distant observations. This makes short-hair black holes an ideal model for studying the physics near singularities. Particularly, in the case of coupling with anisotropic matter fields, short-hair black holes exhibit good stability and satisfy the weak energy condition in physics, providing a new perspective for exploring the extreme behavior of black holes and singularity problems.

In terms of research methods, we first used the WKB method to calculate the QNMs frequencies of short-hair black holes under neutral massless scalar field perturbations. Then, we compared the imaginary part of the QNMs frequency with the surface gravity of the Cauchy horizon and explored whether it complies with the SCCC. Subsequently, we analyzed the perturbative behavior of charged massive scalar fields in short-hair black holes using the WGC to determine under what conditions the black hole can follow the SCCC under scalar field perturbations. The WGC restricts the charge-to-mass ratio of charged massive scalar fields to prevent the formation of naked singularities, thereby supporting the validity of the SCCC in the context of short-hair black holes. By combining the analyses from the WGC and WKB methods, we systematically study the physical behavior of short-hair black holes under extreme conditions and examine whether the SCCC is upheld in the context of short-hair black holes. Through the WKB method analysis, we find that when the black hole's charge $Q$ approaches its maximum value $Q_{\text{max}}$, the SCCC tends to fail. However, as the order $k$ in the metric function $f(r)$ and the angular momentum quantum number $l$ increase, the violation of the SCCC can be mitigated. These results indicate that the magnitude of the black hole's charge $Q$, the angular momentum quantum number $l$, and the order $k$ in the metric function are key factors in studying the physical properties of short-hair black holes and verifying the SCCC. Furthermore, the results from the WGC method indicate that when the charge-to-mass ratio \(\frac{q}{m} > (2k-1)^{\frac{1}{2k}} \), short-hair black holes can satisfy the requirements of the SCCC; however, when \( 2\sqrt{f_{r_0}} (2k-1)^{\frac{1}{2k}} > \frac{q}{m} \), the SCCC will be violated. 

Currently, the WGC method has certain limitations when verifying whether short-hair black holes adhere to the SCCC. Since this method primarily relies on the surface gravity of the event horizon rather than the surface gravity of the Cauchy horizon, it can only provide approximate conditions for the validity of the SCCC, rather than precise conclusions. Particularly in Reissner-Nordström-like black holes, the surface gravity of the Cauchy horizon is generally greater than that of the event horizon, thus limiting the applicability and accuracy of the WGC method in such black holes. 

Future research should focus on developing more precise and broadly applicable methods, especially building theoretical frameworks that account for the effects of both the Cauchy horizon and the event horizon. This would significantly improve the accuracy of SCCC verification and deepen our understanding of the behavior of short-hair black holes within the framework of general relativity.Moreover, future studies should expand to include rotating black holes and other types of black holes, and explore the impact of various effects on SCCC verification. These investigations will contribute to a comprehensive understanding of the SCCC across a broader parameter space. A thorough validation of the SCCC is crucial for preserving the integrity of general relativity. As the fundamental theory describing gravity, the consistency of general relativity is, to some extent, dependent on the validity of the SCCC. If future research can verify the SCCC in a wider range of contexts, it will strengthen the robustness of general relativity and provide new perspectives for exploring black hole physics, gravitational waves, and frontier problems in cosmology.

\section{acknowledgements}
We acknowledge the anonymous referee for a constructive report that has significantly improved this paper.This work was  supported by Guizhou Provincial Basic Research Program(Natural Science)(Grant No.QianKeHeJiChu-[2024]Young166),  the Special Natural Science Fund of Guizhou University (Grant No.X2022133), the National Natural Science Foundation of China (Grant No.12365008) and the Guizhou Provincial Basic Research Program (Natural Science) (Grant No.QianKeHeJiChu-ZK[2024]YiBan027) .

\bibliographystyle{unsrt}
\bibliography{sccc.bib}

\end{document}